\documentclass[conference]{IEEEtran}
\IEEEoverridecommandlockouts
\usepackage{cite}
\usepackage{amsmath,amssymb,amsfonts}
\usepackage{algorithmic}
\usepackage{graphicx}
\usepackage{multirow}
\usepackage{textcomp}

\usepackage{xcolor}
\usepackage{natbib}
\def\BibTeX{{\rm B\kern-.05em{\sc i\kern-.025em b}\kern-.08em
    T\kern-.1667em\lower.7ex\hbox{E}\kern-.125emX}}

\begin{document}

\title{Airline Fleet Assignment Problems with Binary and Integer Programming models: Classical vs Quantum Annealing\\
}
\author{
\IEEEauthorblockN{Kuntal Adak}
\IEEEauthorblockA{\textit{Corporate Incubation} \\
\textit{Tata Consultancy Services}\\
Kolkata, India \\
adak.kuntal@tcs.com}
\and
\IEEEauthorblockN{Sakshi Kaushik}
\IEEEauthorblockA{\textit{Corporate Incubation} \\
\textit{Tata Consultancy Services}\\
Delhi, India \\
sakshi.kaushik@tcs.com}
\and
\IEEEauthorblockN{Rahul Rana}
\IEEEauthorblockA{\textit{Corporate Incubation} \\
\textit{Tata Consultancy Services}\\
Kolkata, India \\
rahul.rana2@tcs.com}
}

\maketitle

\begin{abstract}
The fleet assignment problem (FAP) in the context of airline operations is to optimally assign different types of aircraft (fleet types) to scheduled flights in a way that minimizes the operational cost, accounting for each aircraft's capacity and the demand on each flight, while adhering to a set of operational constraints. This study explores two distinct mathematical formulations of the FAP: a Binary Linear Programming (BLP) model and an Integer Linear Programming (ILP) model. These models are addressed using both SCIP (Solving Constraint Integer Programs) and Quantum Annealers to assess their efficacy in solving this complex problem. The integer model incorporates aircraft balance constraint to track the number of aircraft at a particular location at the end of the day to reuse the aircraft efficiently. We demonstrated the application of the model through a set of problem instances generated based on one of the Australia’s airline data, highlighting the benefits of effective fleet assignment in improving fleet utilization, reducing costs, and enhancing overall operational efficiency. Our experiments reveal that for the BLP model, quantum annealers and SCIP provide identical optimal solutions for smaller instances of the problem. However, as the problem size increases, quantum annealers consistently deliver near-optimal solutions with reduced computational time compared to SCIP. In contrast, for the ILP model, quantum annealers achieve optimal solutions for smaller instances but struggle to find feasible solutions for larger ones. SCIP, on the other hand, remains effective in managing larger instances, though with increased computational effort. Additionally, we discuss the scalability of the solution approach and its relevance to real-time airline scheduling systems. This research highlights the potential of quantum annealing in tackling large-scale optimization problems within the airline industry, demonstrating its efficiency for certain problem sizes while also acknowledging its current limitations. The comparative analysis provides valuable insights into the performance of advanced computational techniques, paving the way for further advancements in optimizing fleet assignments in the aviation sector.
\end{abstract}


\section{Introduction}
Airline scheduling is a complex and multifaceted problem involving the optimization of various interconnected components that include flight scheduling, fleet assignment, crew scheduling, aircraft routing, and maintenance scheduling \cite{bib2}. The fleet assignment problem (FAP) is a critical component of airline operations and plays an important role in the broader context of airline scheduling. It involves determining the optimal assignment of various aircraft types from an airline’s fleet to a set of scheduled flights, with the goal of optimizing key performance metrics such as cost, revenue, fleet utilization, and operational efficiency \cite{bib1}. 

Airlines typically operate diverse fleets, with each aircraft type offering different capacities, ranges, fuel efficiencies, and operational costs. The challenge is matching the right aircraft type to the right flight, considering passenger demand and operational constraints. A well-executed fleet assignment ensures that the airline's schedule is both operationally feasible and economically viable, setting the stage for efficient day-to-day operations. It provides the foundation for subsequent scheduling activities, such as crew assignment, maintenance planning, and real-time operational adjustments.  

The predefined fleet allocation simplifies the task of assigning individual aircraft, as the tail assignment process only needs to match the right aircraft within each fleet type to the flights rather than determining the best type of aircraft for each route from scratch. It ensures that the required number of aircraft types is available at each airport. This balanced distribution of fleet types facilitates a smoother tail assignment process because it minimizes the need for repositioning aircraft and reduces the risk of having insufficient aircraft available for scheduled flights. Fleet assignment helps optimize the use of different aircraft types based on their characteristics (e.g., range, capacity, fuel efficiency). A well-executed fleet assignment reduces the likelihood of scheduling conflicts and operational disruptions during tail assignment, leading to better on-time performance and reduced costs associated with delays or cancellations. Fleet assignment provides valuable data that can be used in the tail assignment stage to make informed decisions. For example, historical data on how certain fleet types perform on specific routes can guide the assignment of individual aircraft, optimizing for reliability, fuel efficiency, and passenger comfort. 

This assignment problem is inherently complex due to the multitude of variables and constraints involved. Airlines typically operate a mix of aircraft types, each with unique operating costs and capacities. 

Fleet assignment has been tackled using mathematical optimization techniques such as \textbf{Binary Linear Programming (BLP)} and \textbf{Integer Linear Programming (ILP)} \cite{bib1}. These formulations allow airlines to balance operational costs with other factors, such as fleet availability and passenger demand. While classical optimization solvers like \textbf{SCIP (Solving Constraint Integer Programs)} \cite{bib18} have been widely used for this purpose, the growing scale and complexity of real-world problems pose significant computational challenges. As airlines expand their fleets and schedules, the need for more scalable and efficient solution approaches becomes paramount. 

Recent advances in quantum computing, particularly \textbf{Quantum Annealing}, offer a promising alternative for solving large-scale optimization problems \cite{bib13}. Quantum annealers have demonstrated the ability to generate high-quality solutions with faster computational times, especially for problems that become increasingly difficult or intractable for classical approaches. Hybrid Quantum Annealing solvers leverage the complementary strengths of classical and quantum resources, with classical algorithms guiding the problem-solving process while quantum computing explores the solution space more efficiently. \textbf{Constrained Quadratic Models (CQMs)} allow for the inclusion of constraints directly in the model, offering a more flexible and expressive framework for real-world optimization problems \cite{bib12}. In this study, we explore the application of quantum annealing alongside SCIP to solve two distinct mathematical formulations of FAP—BLP and ILP. 

The paper is organized as follows. In Section \uppercase\expandafter{\romannumeral 2\relax}, we present the Literature review highlighting the earlier related works done. In Section \uppercase\expandafter{\romannumeral 3\relax}, we present the general mathematical model (BLP and ILP), objective function of the mathematical model for the fleet assignment problem. Section \uppercase\expandafter{\romannumeral 4\relax} discusses the methodology and solution approaches. Then we discuss the results of the fleet assignment problem in section \uppercase\expandafter{\romannumeral 5\relax} and conclusions and future scope in section \uppercase\expandafter{\romannumeral 6\relax} \& \uppercase\expandafter{\romannumeral 7\relax} respectively.

\begin{figure}[!h]
\centering
\includegraphics[scale=0.5]{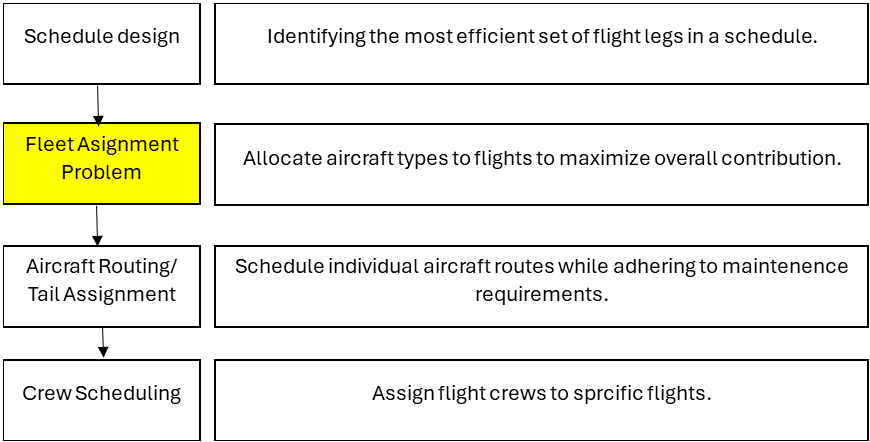}
\caption{Flight scheduling Process overview}
\label{fig:flow}
\end{figure}

\section{Literature Survey}

Optimization of Fleet Assignment: A Case Study in Turkey \cite{bib1} proposed an integer model for fleet assignment problem. They seek to reduce costs including both operational and passenger spill cost, and to optimize fleets. 

Airline fleet assignment concepts, models, and algorithms \cite{bib2} offer a comprehensive tutorial on the foundational and advanced models and methods developed for the Fleet Assignment Problem (FAP), including its integration with other airline decision-making processes. 

Robust Airline Fleet Assignment: Station Purity Enforcement Using Station Decomposition \cite{bib3} offers a strong solution by restricting the number of fleet types permitted to operate at each airport in the schedule. 

Advances in the Optimization of Airline Fleet Assignment \cite{bib4} model the problem as mixed-integer multicommodity flow on networks encoding activities linking flight departures which permits a profitability-based tradeoff between operational goals and revenue 

Mathematical models in airline schedule planning: A survey \cite{bib5} survey these various models and solution techniques in the recent past for almost all of the problems related to fleet assignment.  

Applying Integer Linear Programming to the Fleet Assignment Problem \cite{bib6} formulated and solved the fleet assignment problem as an integer linear programming model, permitting assignment of two or more fleets to a flight schedule simultaneously. 

Approaches to Incorporating Robustness into Airline Scheduling \cite{bib7} demonstrates how to generate multiple optimal solutions for the Aircraft Routing problem and proposes ways to assess the robustness of these solutions. 

Itinerary based- Airline fleet assignment \cite{bib8} presents a new formulation and solution method that incorporates network effects, with the goal of maximizing profits when assigning aircraft types to flight legs. 

Airline fleet assignment: a state of the art \cite{bib9} is an adaptation of the latest advancements in fleet assignment models from the Operations Research literature. The various models are evaluated based on their applicability to real-world problems in a short-term operational context, while also considering related issues such as maintenance planning and crew assignment. 

Applying Quantum Annealing to the Tail Assignment Problem \cite{bib10} where the problem is formulated as a Quadratic Unconstrained Binary Optimization (QUBO) model, approached using various techniques, and solved with both a classical solver and two hybrid solvers. 

Applying the Quantum Approximate Optimization Algorithm to the Tail-Assignment Problem \cite{bib11} simulate the quantum approximate optimization algorithm (QAOA) applied to instances of this problem derived from real-world data.

\section{Problem Formulation}
The fleet assignment problem is expressed as an optimization model, utilizing both a Binary Linear Programming (BLP) model and an Integer Linear Programming (ILP) model, incorporating the following key components: 

\subsection{BLP Model for Multi-Day Fleet Type Assignment:}
\textbf{Parameters}

\begin{itemize}
  \item $Q_j$ : Capacity of fleet type j
  \item $D_{id}$ : Demand or forecasted seats for flight i on day d
  \item $C_{ijd}$ : Operational Cost for flight i corresponding to fleet j on day d
  \item $N_j$ : Maximum number of available aircraft of type j
  \item $D$ : Set of days
  \item $F_d$ : Set of flights at day d
  \item $K$  : Set of fleet types
\end{itemize}

\textbf{Decision variables}

\begin{itemize}
  \item $x_{ijd} = 1$ : if fleet type j is assigned to flight i on day d, and 0 otherwise. 
\end{itemize}

\textbf{Objective function}

The objective is to minimize the operational cost, accounting for each aircraft's capacity and the demand on each flight, while adhering to a set of operational constraints. This can be expressed as:
\begin{equation}\label{eq:1}
min \sum_{d \in D, i \in F_{d}, j \in K} C_{ijd} \* x_{ijd} + \lambda \sum_{d \in D, i \in F_{d}, j \in K} [Q_j - D_{id}]^2 x_{ijd}
\end{equation}
Where, $\lambda \in \mathbb{R}^+$ represents the penalty for violating constraint accounting for the difference between each aircraft's capacity and the demand on each flight.

\textbf{Constraints}
\begin{enumerate}
  \item Each flight is allocated exactly one fleet type:
    \begin{equation}\label{eq:2}
    \sum_{j \in K} x_{ijd} = 1, \forall i \in F_{d}, \forall d \in D
    \end{equation}
  \item The number of aircraft assigned does not exceed the maximum available aircrafts:
    \begin{equation}\label{eq:3}
    \sum_{i \in F_d} x_{ijd} \leq N_j, \forall j \in K, \forall d \in D
    \end{equation}  
\end{enumerate}

\textbf{Search Space Analysis}

Let, $\kappa_d = |F_d|=$ number of flights on day $d$.
$\eta = $Number of fleet-types.
For each day, $\eta$ aircraft types can be assigned to each of the $\kappa_d$ flights. So, the search space for each day is 
$\eta^{\kappa_d}$.

For $D$ days, the overall search space size is the product of the search spaces for each individual day. Therefore, the total size of the search space can be represented as: 
\begin{equation}\label{eq:4}
 \prod _{d\in D} \eta ^{\kappa _d} = \eta ^{\sum _{d\in D} \kappa _d} \approx O\left(\eta ^M\right), where \; M = \sum _{d\in D}\kappa _d
\end{equation}

Thus, the search space for the Fleet Type Assignment Problem (FTAP) in a multi-day operation scenario can be succinctly represented as $O\left(\eta ^M\right)$, where $\eta$ is the number of fleet types and $M$ is the total number of flights across all days. This highlights the exponential growth in complexity as either the number of aircraft types or the total number of flights increases, which makes this problem intractable in nature.
The actual feasible search space may be smaller due to constraints Eq.\ref{eq:1}, Eq.\ref{eq:2}
, but the upper bound of the search space remains 
$O\left(\eta ^M\right)$.

Classical solvers such as SCIP will find it challenging to address complex problems like the FTAP within polynomial time, as the search space expands exponentially with the size of the problem.

\subsection{ILP model for Single-Day Fleet Type Assignment}
\textbf{Parameters}
\begin{itemize}
    \item $Q_j :$ Capacity of fleet type $j$
    \item $D_i :$ Demand or forecasted seats for flight $i$
    \item $C_{ij} :$ Operational Cost for flight $i$ corresponding to fleet type $j$
    \item $N_j :$ Maximum number of available aircraft of type $j$
    \item $F :$ Set of flights
    \item $K :$ Set of fleets
    \item $M :$ Set of Nodes
    \item $a_{ik}=1$ if flight $i$ is an arrival at node $k$, else 0
    \item $b_{ik}=1$ if flight $i$ is a departure at node $k$, else 0 
\end{itemize}

\textbf{Decision variables}
\begin{itemize}
    \item $x_{ij}=1$, if fleet type $j$ is assigned to flight $i$
    \item $G_{kj} :$ Number of aircrafts of fleet type $j$ grounded at node $k$. $G_{kj} \in \mathbb{Z}^+$
\end{itemize}

\textbf{Objective function}
\begin{equation}\label{eq:5}
 min \sum_{i \in F, j \in K} C_{ij} x_{ij} + \lambda \sum_{i \in F, j \in K} [Q_j - D_i]^2 x_{ij}, where \; \lambda \in \mathbb{R}^+
\end{equation}

\textbf{Constraints}
\begin{enumerate}
  \item Each flight is allocated exactly one fleet type:
    \begin{equation}\label{eq:6}
    \sum_{j \in K} x_{ij} = 1, \forall i \in F
    \end{equation}
  \item The number of aircraft assigned does not exceed the maximum available aircrafts:
    \begin{equation}\label{eq:7}
    \sum_{i \in F} x_{ij} \leq N_j, \forall j \in K
    \end{equation}  
  \item Aircraft balance constraints. The number of aircraft of a given fleet type at any node $(G_{kj})$ is calculated by taking the number of aircraft of that type just before the node $(G_{k-1j})$, adding the 
arrivals $(a_{ik})$ and subtracting the departures $(b_{ik})$.
    \begin{equation}\label{eq:8}
    G_{k-1j} + \sum _{i\in F}(a_{ik}+b_{ik})x_{ij} = G_{kj} ,\forall k\in M, \forall j\in K
    \end{equation}
\end{enumerate}

\section{Methodology }
\subsection{Data}
We consider a real-world airline network utilizing one of the Australian airline data with 46, 92, 184, 276, 368, 1104, 1840, 3680 flights per day and a fleet of 4 different aircraft types, each with varying capacities and operating costs.

Following is the sample input dataset: 

\begin{table}[!ht]
\centering
\caption{Sample Fleet data}
\begin{tabular}{|c|c|c|}
\hline
\textbf{Fleet} & \textbf{Capacity} & \textbf{Number of Aircrafts} \\
\hline
A330 & 159 & 10 \\
\hline
A220 & 192 & 15 \\
\hline
B737 & 142 & 15 \\
\hline
B717 & 165 & 8 \\
\hline
\end{tabular}
\label{tab:fleet_data}
\end{table}

\begin{table}[!ht]
\centering
\caption{Sample Flight Schedule}
\resizebox{\columnwidth}{!}{%
\begin{tabular}{|c|c|c|c|c|c|c|}
\hline
\textbf{Flight Number} & \textbf{Origin} & \textbf{Departure Time} & \textbf{Destination} & \textbf{Arrival Time} & \textbf{Number of Passengers} & \textbf{Day} \\
\hline
11111 & SYD & 6:15:00 & MEL & 7:20:00 & 157 & 1 \\
\hline
11112 & HBA & 7:00:00 & MEL & 8:50:00 & 207 & 1 \\
\hline
11113 & SYD & 7:30:00 & MEL & 8:35:00 & 147 & 1 \\
\hline
11114 & MEL & 7:35:00 & OOL & 9:10:00 & 113 & 2 \\
\hline
11115 & DRW & 8:25:00 & MEL & 9:40:00 & 190 & 2 \\
\hline
11116 & MEL & 8:30:00 & ADA & 10:00:00 & 141 & 2 \\
\hline
11117 & SYD & 9:00:00 & MEL & 10:05:00 & 157 & 3 \\
\hline
11118 & MEL & 9:00:00 & SYD & 10:05:00 & 145 & 3 \\
\hline
11119 & SYD & 10:00:00 & MEL & 11:05:00 & 159 & 3 \\
\hline
\end{tabular}%
}
\label{tab:flight_schedule}
\end{table}

\begin{table}[!ht]
\centering
\caption{Sample Flight-Fleet-Cost Data}
\resizebox{\columnwidth}{!}{
\begin{tabular}{|c|c|c|c|c|c|c|}
\hline
\textbf{Flight Number} & \textbf{From} & \textbf{To} & \textbf{Fleet Type} & \textbf{Crew Cost (AUD)} & \textbf{Operational Cost (AUD)} & \textbf{Total Cost (AUD)} \\
\hline
11111 & SYD & MEL & A330 & 1945.15 & 2745.25 & 4690.40 \\
\hline
11112 & HBA & MEL & A330 & 1945.15 & 2745.25 & 5740.80 \\
\hline
11113 & SYD & MEL & A330 & 817.80 & 1818.20 & 6266.00 \\
\hline
11114 & MEL & OOL & A330 & 1945.15 & 2745.25 & 5740.80 \\
\hline
11115 & DRW & MEL & A330 & 2765.00 & 2450.00 & 5215.60 \\
\hline
11116 & MEL & ADA & A330 & 1945.15 & 2745.25 & 5740.80 \\
\hline
11117 & SYD & MEL & A330 & 2765.00 & 2450.00 & 5215.60 \\
\hline
11118 & MEL & SYD & A330 & 1945.15 & 2745.25 & 4690.40 \\
\hline
11119 & SYD & MEL & A330 & 2765.00 & 2450.00 & 5215.60 \\
\hline
\end{tabular}}

\label{tab:flight_fleet_cost}
\end{table}

\subsection{ Solution Approach }
The fleet assignment problem can be modeled as a Binary Linear Programming (BLP) model and an Integer Linear Programming (ILP) model. The integer model incorporates aircraft balance constraint to track the number of aircraft at a particular location at the end of the day to reuse the aircraft efficiently. We employ an exact method using commercial optimization solvers such as SCIP and quantum annealer based hybrid approach. 
\subsection{ Solver}
\subsubsection{SCIP}
SCIP (Solving Constraint Integer Programs) relies on LP solvers (such as SoPlex, CPLEX, or Gurobi) to solve the linear relaxation of the optimization problem \cite{bib18}. LP relaxations play a crucial role in solving MIP problems by providing bounds and guiding the search process in branch-and-bound.

\subsubsection{Quantum Annealing}
Quantum annealing utilizes quantum tunneling, enabling it to overcome local minima and potentially discover superior (lower cost) solutions compared to classical methods \cite{bib15}. The annealing process seeks the lowest energy configuration, which corresponds to the optimal solution. In many cases, quantum annealers are combined with classical post-processing techniques to fine-tune and verify solutions. This hybrid approach can help correct minor violations of constraints or further refine the optimal solution \cite{bib16}. Hybrid solvers leverage the use of both classical and quantum resources to solve problems while maximizing their complementary capabilities. These solvers are designed to handle large-scale, real-world problems by utilizing classical algorithms to guide the problem-solving process and quantum computing to efficiently explore the solution space. The hybrid workflow methodology is designed to address the shortcomings of the fully quantum approach by utilizing multiple resolution paths that incorporate both quantum and classical techniques. Typically, these systems operate by first using a classical solver to handle the problem (whether it's BQM, DQM, or CQM). Following this, one or more hybrid heuristic solvers are initiated to run concurrently, searching for solutions. Each heuristic solver comprises a classical heuristic model that investigates the solution space. Quantum annealers, such as D-Wave’s systems, are built to solve QUBO problems, and CQMs can be solved on these devices by encoding them into Quadratic Unconstrained Binary Optimization (QUBO). 
\begin{figure}[!h]
\centering
\includegraphics[scale=0.3]{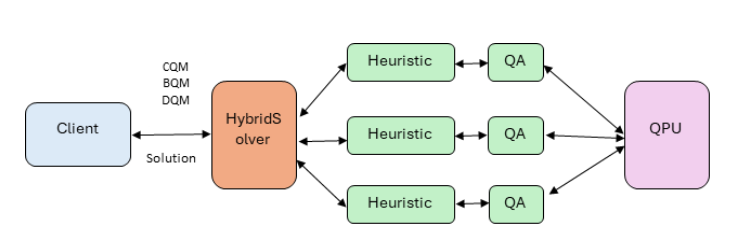}
\caption{Architecture of Quantum Hybrid Solvers \cite{bib19}}
\label{fig:archi_Qhybrid}
\end{figure}

The problem is formulated using a Constrained Quadratic Model (CQM), which offers greater expressive capabilities compared to Binary Quadratic Models (BQMs), as it allows the inclusion of constraints and the use of variables beyond binary. A CQM is a generalization of the widely used QUBO model, commonly applied in quantum annealing. Unlike QUBOs, CQMs can directly incorporate constraints, enabling more detailed representations of real-world optimization problems. In this case, we are utilizing the Leap Hybrid CQM solver (LHCQM) to address the problem described in this paper.

\section{Results}
\subsection{BLP Model for Multi-Day Fleet Type Assignment: }
\begin{figure}[!h]
\centering
\includegraphics[scale=0.55]{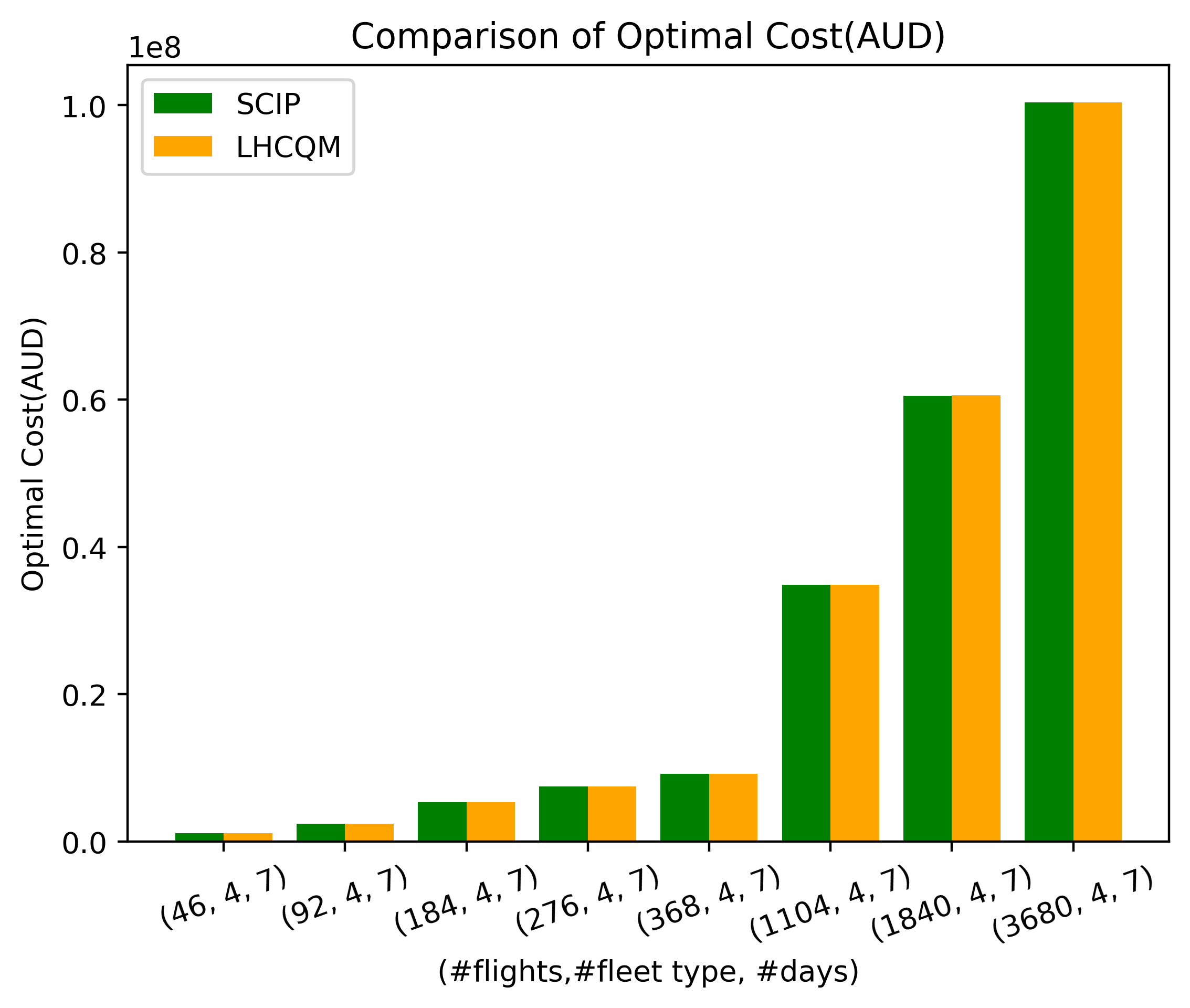}
\caption{Comparison of optimal cost (AUD) for both solvers in Fleet assignment (BLP) model}
\label{fig:blp_optcost}
\end{figure}

The following table compares the performance of SCIP and quantum annealing based hybrid solver in terms of Optimal Cost and Execution Time (seconds) for solving the Fleet Assignment Problem (BLP model) across various flight sizes. The table highlights the results as the number of flights increases.  Notably, the SCIP solver successfully identified the optimal solution for each instance of the problem in our experiment, although it required a substantial amount of time to solve the larger instances (with daily flight counts of 1104, 1840, and 3680) [Table \ref{tab:blp_optimalcost_exetime}]. In contrast, LHCQM solver is able to produce the optimal solution till problem size with 92 flights [Table \ref{tab:blp_optimalcost_exetime}] and near-optimal solutions up to a flight size of 3680, all with significantly shorter solving times [figure \ref{fig:blp_optcost}] compared to SCIP. 

The comparison reflects how computational complexity affects both cost minimization and execution time as the problem scales.

\begin{table}[!ht]
\centering
\caption{Result Comparison table for Fleet Assignment model (BLP) for Optimal Cost and Execution time (sec)}
\resizebox{\columnwidth}{!}{%
\begin{tabular}{|c|c|cc|cc|}
\hline
\multirow{2}{*}{\textbf{Flights, fleet, day}} & \multirow{2}{*}{\textbf{Variables, Constraints}} & \multicolumn{2}{c|}{\textbf{Optimal cost (AUD)}} & \multicolumn{2}{c|}{\textbf{Execution time (sec)}} \\ \cline{3-6} 
                                              &                                                  & \multicolumn{1}{c|}{SCIP}         & LHCQM        & \multicolumn{1}{c|}{SCIP}           & LHCQM        \\ \hline
(46,4, 7)                                     & (1288, 350)                                      & \multicolumn{1}{c|}{1151514}      & 1151514      & \multicolumn{1}{c|}{1.3}            & 2.2          \\ \hline
(92,4, 7)                                     & (2576, 672)                                      & \multicolumn{1}{c|}{2424749.6}    & 2424749.6    & \multicolumn{1}{c|}{2.7}            & 2.8          \\ \hline
(184, 4, 7)                                   & (5152, 1316)                                     & \multicolumn{1}{c|}{5364340.8}    & 5365682.4    & \multicolumn{1}{c|}{7.2}            & 7.7          \\ \hline
(276, 4, 7)                                   & (7728, 1960)                                     & \multicolumn{1}{c|}{7470153.6}    & 7488395.2    & \multicolumn{1}{c|}{13.4}           & 14.1         \\ \hline
(368, 4, 7)                                   & (10304, 2604)                                    & \multicolumn{1}{c|}{9146992.4}    & 9147517.6    & \multicolumn{1}{c|}{24.9}           & 24.8         \\ \hline
(1104,4, 7)                                   & (30912, 7756)                                    & \multicolumn{1}{c|}{34821682}     & 34862649.2   & \multicolumn{1}{c|}{233.8}          & 153.2        \\ \hline
(1840, 4, 7)                                  & (51520, 12908)                                   & \multicolumn{1}{c|}{60505526.2}   & 60627604.6   & \multicolumn{1}{c|}{1138.0}         & 216.0        \\ \hline
(3680, 4, 7)                                  & (103040, 25788)                                  & \multicolumn{1}{c|}{100393602.4}  & 100395635.6  & \multicolumn{1}{c|}{11579.2}        & 976.7        \\ \cline{1-6} 
\end{tabular}%
}
\label{tab:blp_optimalcost_exetime}
\end{table}

\begin{table}[!ht]
\centering
\caption{Sample result getting from SCIP and LHCQM for Fleet Assigned to each schedule flights (problem size: 46 flights per day)}
\resizebox{\columnwidth}{!}{%
\begin{tabular}{|c|c|c|c|c|c|c|c|}
\hline
\textbf{Flights} & \textbf{Origin} & \textbf{Departure} & \textbf{Destination} & \textbf{Arrival} & \textbf{No of Passengers} & \textbf{Day No} & \textbf{Fleet Assigned} \\
\hline
11111            & SYD             & 06:15:00           & MEL                  & 07:20:00         & 157                       & 1               & A330                     \\
\hline
11112            & HBA             & 07:00:00           & MEL                  & 08:50:00         & 207                       & 1               & A220                     \\
\hline
11113            & SYD             & 07:30:00           & MEL                  & 11:15:00         & 147                       & 1               & B717                     \\
\hline
...              & ...             & ...                & ...                  & ...              & ...                       & ...             & ...                      \\
\hline
11156            & DRW             & 21:10:00           & MEL                  & 22:25:00         & 136                       & 1               & A220                     \\
\hline
...              & ...             & ...                & ...                  & ...              & ...                       & ...             & ...                      \\
\hline
111390           & MEL             & 07:35:00           & OOL                  & 09:10:00         & 113                       & 7               & B737                     \\
\hline
11391            & DRW             & 08:25:00           & MEL                  & 09:40:00         & 190                       & 7               & A220                     \\
\hline
111392           & MEL             & 08:30:00           & ADA                  & 10:00:00         & 141                       & 7               & B737                     \\
\hline
...              & ...             & ...                & ...                  & ...              & ...                       & ...             & ...                      \\
\hline
111431           & DRW             & 21:10:00           & MEL                  & 22:25:00         & 131                       & 7               & A220 \\
\hline
\end{tabular}%
}
\label{tab:blp_assgnfleet_scip}
\end{table}

\begin{figure}[!h]
\centering
\includegraphics[scale=0.33]{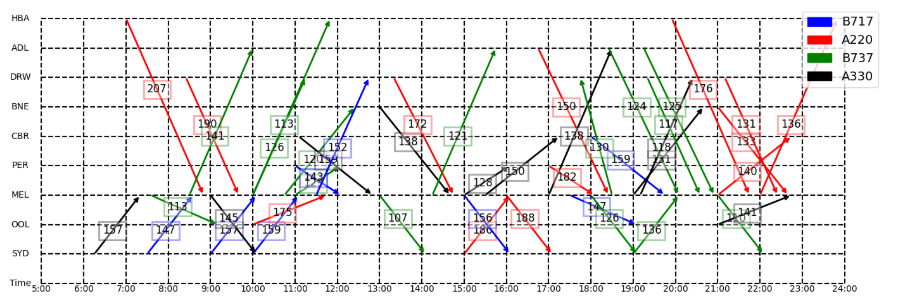}
\caption{Flight to Fleet assignment (BLP) Time-line network (for the day1 results getting from SCIP and LHCQM) depicting the flights departing and arriving at a particular location. Each fleet type is depicted with a different color. Flights are assigned based on the objective of minimizing the cost, considering the difference between capacity and demand of passengers while meeting a set of operational and logistical constraints.  The incoming arrow shows the flights arriving at a location and the outgoing arrows show the flights departing from that location}
\label{fig:blp_assgnflight}
\end{figure}

\begin{figure}[!h]
\centering
\includegraphics[scale=0.55]{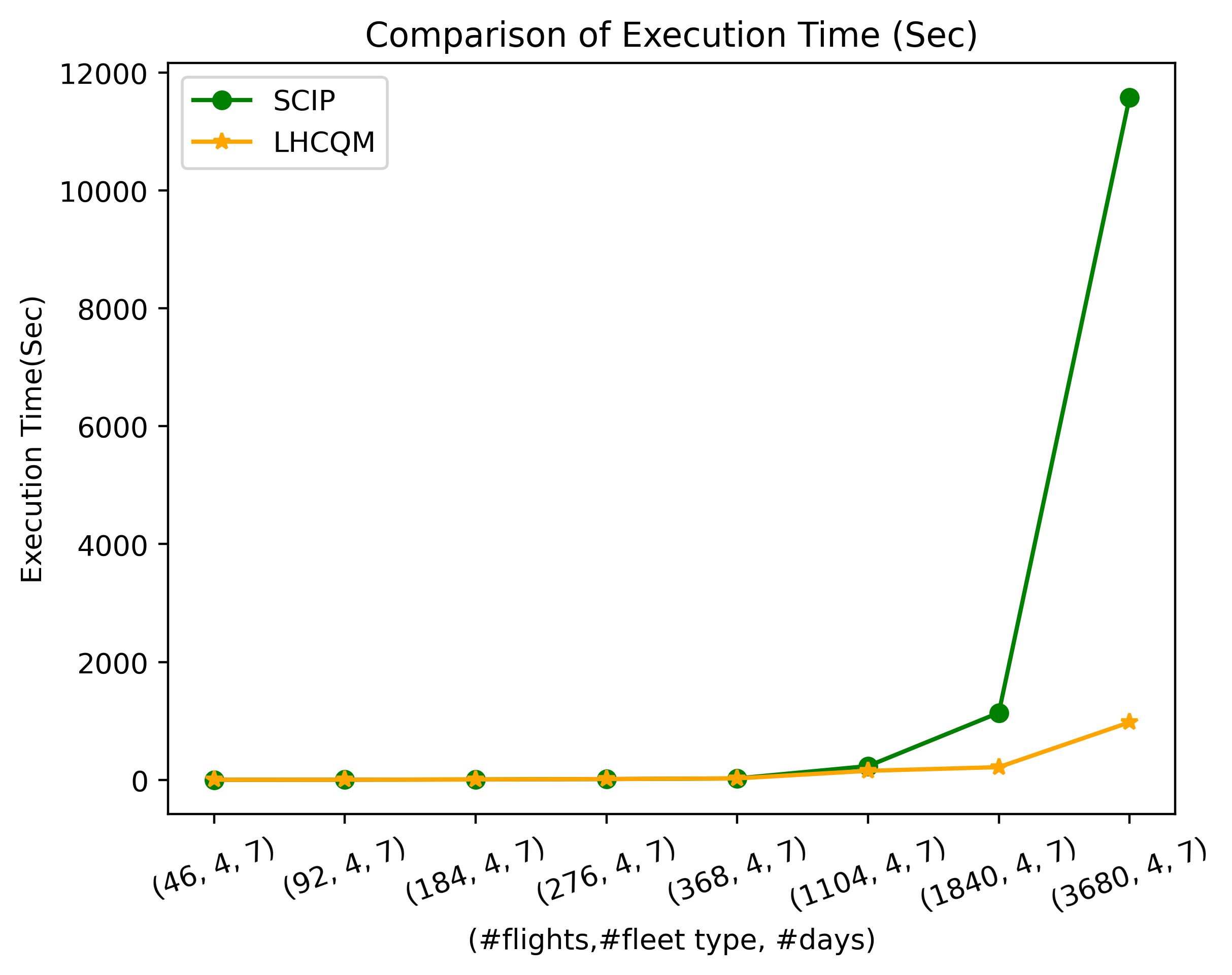}
\caption{Comparison of solution time (sec) for both solvers in Fleet assignment (BLP) model}
\label{fig:blp_soltime}
\end{figure}

\newpage

\subsection{ILP Model for Single-Day Fleet Type Assignment:}

The table below compares the performance of various solvers in terms of Optimal Cost and Execution Time (in seconds) when addressing the Fleet Assignment Problem (ILP model) across different flight sizes. It showcases the results as the number of flights increases. Notably, the LHCQM solver can generate feasible and near-optimal solutions for flight sizes up to 368, while other solvers are employed to find optimal solutions for larger problem sizes. This comparison illustrates how computational complexity impacts both cost minimization and execution time as the problem size grows. 
\begin{table}[!ht]
\centering
\caption{Result Comparison table for Fleet Assignment model (ILP) for Optimal Cost and Execution time(sec)}
\resizebox{\columnwidth}{!}{%
\begin{tabular}{|c|c|cc|cc|}
\hline
\multirow{2}{*}{\textbf{\begin{tabular}[c]{@{}c@{}}Flights,\\  fleet\end{tabular}}} & \multirow{2}{*}{\textbf{\begin{tabular}[c]{@{}c@{}}Variables, \\ Constraints\end{tabular}}} & \multicolumn{2}{c|}{\textbf{Optimal cost (AUD)}} & \multicolumn{2}{c|}{\textbf{Execution time (sec)}} \\ \cline{3-6} 
                                                                                    &                                                                                             & \multicolumn{1}{c|}{SCIP}         & LHCQM        & \multicolumn{1}{c|}{SCIP}           & LHCQM        \\ \hline
(46,4)                                                                              & (536, 402)                                                                                  & \multicolumn{1}{c|}{121274.4}     & 121274.4     & \multicolumn{1}{c|}{1.6653}         & 1.6693       \\ \hline
(92,4)                                                                              & (1104, 832)                                                                                 & \multicolumn{1}{c|}{242548.8}     & 242548.8     & \multicolumn{1}{c|}{3.0952}         & 3.6358       \\ \hline
(184, 4)                                                                            & (1956, 1408)                                                                                & \multicolumn{1}{c|}{485097.6}     & 528372.0     & \multicolumn{1}{c|}{5.2123}         & 5.9468       \\ \hline
(276, 4)                                                                            & (3044, 2220)                                                                                & \multicolumn{1}{c|}{727646.4}     & 810180.8     & \multicolumn{1}{c|}{8.2135}         & 9.7220       \\ \hline
(368, 4)                                                                            & (3912, 2812)                                                                                & \multicolumn{1}{c|}{970195.2}     & 1075896.0    & \multicolumn{1}{c|}{10.6262}        & 11.8141      \\ \hline
(1104,4)                                                                            & (12744, 9436)                                                                               & \multicolumn{1}{c|}{2910586.0}    & Infeasible   & \multicolumn{1}{c|}{36.8492}        & NA           \\ \hline
(1840, 4)                                                                           & (21576, 16060)                                                                              & \multicolumn{1}{c|}{4850976.0}    & Infeasible   & \multicolumn{1}{c|}{67.8877}        & NA           \\ \hline
(3680, 4)                                                                           & (43656, 32620)                                                                              & \multicolumn{1}{c|}{9701952.0}    & Infeasible   & \multicolumn{1}{c|}{173.1200}       & NA           \\ \cline{1-6} 
\end{tabular}%
}
\label{tab:ilp_optimalcost_exetime}
\end{table}

\begin{table}[!ht]
\centering
\caption{Sample result for Fleet Assignment model (ILP) for SCIP (problem size: 46 flights per day) }
\resizebox{\columnwidth}{!}{%
\begin{tabular}{|c|c|c|c|c|c|c|c|}
\hline
\textbf{Flight} & \textbf{Origin} & \textbf{Destination} & \textbf{Departure time} & \textbf{Arrival time} & \textbf{Assigned Fleet Type} & \textbf{Number of Passengers} 
 \\
\hline
11111 & SYD & MEL & 10:00:00 & 11:05:00 & A330 & 159 \\
\hline
11112 & OOL & MEL & 10:00:00 & 11:45:00 & A330 & 175 \\
\hline
11113 & MEL & DRW & 10:00:00 & 11:15:00 & B737 & 126 \\
\hline
11114 & MEL & HBA & 10:00:00 & 11:50:00 & B737 & 113 \\
\hline
11115 & MEL & HBA & 10:00:00 & 11:50:00 & B717 & 136 \\
\hline
11116 & MEL & BNE & 10:45:00 & 12:25:00 & B737 & 120 \\
\hline
11117 & MEL & PER & 11:00:00 & 12:05:00 & A330  & 129 \\
\hline
11118 & PER & MEL & 11:00:00 & 12:05:00 & B737 & 143 \\
\hline
11119 & CBR & MEL & 11:05:00 & 12:50:00 & B737 & 159 \\
\hline
11120 & MEL & DRW  & 11:30:00 & 12:45:00 & B717 & 152 \\
\hline
11121 & BNE & MEL & 1:00:00 & 2:40:00 & B737 & 138 \\
\hline
11122 & MEL & SYD & 1:00:00 & 2:05:00 & B737 & 107 \\
\hline
11123 & DRW & MEL & 1:20:00 & 2:45:00 & A330 & 172 \\
\hline
11124 & MEL & ADA & 2:15:00 & 3:45:00 & B737 & 121 \\
\hline
\end{tabular}%
}
\label{tab:ilp_assgnfleet_scip}
\end{table}

\begin{table}[!ht]
\centering
\caption{Optimal number of grounded aircraft at each airport (SCIP) (problem size: 46 flights per day)}
\begin{tabular}{|c|c|c|c|c|}
\hline
\textbf{City} & \textbf{A220} & \textbf{A330} & \textbf{B717} & \textbf{B737}
 \\
\hline
ADA & 0 & 0 & 0 & 0 \\ \hline
SYD & 1 & 1 & 0 & 0 \\ \hline
DRW & 0 & 1 & 0 & 0 \\ \hline
BNE & 0 & 0 & 0 & 0 \\ \hline
OOL & 0 & 1 & 0 & 1 \\ \hline
CBR & 0 & 0 & 0 & 0 \\ \hline
MEL & 1 & 0 & 2 & 3 \\ \hline
PER & 0 & 0 & 0 & 1 \\ \hline
HBA & 0 & 0 & 0 & 0 \\
\hline
\end{tabular}
\label{tab:ilp_groundaircraft_scip}
\end{table}

\begin{table}[!ht]
\centering
\caption{Sample result for Fleet Assignment model (ILP) for LHCQM (problem size: 46 flights per day)}
\resizebox{\columnwidth}{!}{%
\begin{tabular}{|c|c|c|c|c|c|c|c|}
\hline
\textbf{Flight} & \textbf{Origin} & \textbf{Destination} & \textbf{Departure time} & \textbf{Arrival time} & \textbf{Assigned Fleet Type} & \textbf{Number of Passengers} 
 \\
\hline
11111 & SYD & MEL & 10:00:00 & 11:05:00 & A330 & 159 \\
\hline
11112 & OOL & MEL & 10:00:00 & 11:45:00 & B737 & 175 \\
\hline
11113 & MEL & DRW & 10:00:00 & 11:15:00 & B737 & 126 \\
\hline
11114 & MEL & HBA & 10:00:00 & 11:50:00 & B717 & 113 \\
\hline
11115 & MEL & HBA & 10:00:00 & 11:50:00 & B737 & 136 \\
\hline
11116 & MEL & BNE & 10:45:00 & 12:25:00 & A330 & 120 \\
\hline
11117 & MEL & PER & 11:00:00 & 12:05:00 & A220  & 129 \\
\hline
11118 & PER & MEL & 11:00:00 & 12:05:00 & B717 & 143 \\
\hline
11119 & CBR & MEL & 11:05:00 & 12:50:00 & A330 & 159 \\
\hline
11120 & MEL & DRW  & 11:30:00 & 12:45:00 & A220 & 152 \\
\hline
11121 & BNE & MEL & 1:00:00 & 2:40:00 & B717 & 138 \\
\hline
11122 & MEL & SYD & 1:00:00 & 2:05:00 & B737 & 107 \\
\hline
11123 & DRW & MEL & 1:20:00 & 2:45:00 & B737 & 172 \\
\hline
11124 & MEL & ADA & 2:15:00 & 3:45:00 & A330 & 121 \\
\hline
\end{tabular}%
}
\label{tab:ilp_assgnfleet_scip}
\end{table}

\begin{table}[!ht]
\centering
\caption{Optimal number of grounded aircraft at each airport (LHCQM) (problem size: 46 flights per day)}
\begin{tabular}{|c|c|c|c|c|}
\hline
\textbf{City} & \textbf{A220} & \textbf{A330} & \textbf{B717} & \textbf{B737}
 \\
\hline
ADA & 2 & 1 & 2 & 0 \\ \hline
SYD & 1 & 2 & 6 & 0 \\ \hline
DRW & 0 & 0 & 1 & 0 \\ \hline
BNE & 0 & 2 & 1 & 1 \\ \hline
OOL & 2 & 0 & 0 & 3 \\ \hline
CBR & 2 & 4 & 0 & 2 \\ \hline
MEL & 17 & 9 & 8 & 5 \\ \hline
PER & 0 & 1 & 0 & 3 \\ \hline
HBA & 0 & 1 & 0 & 0 \\
\hline
\end{tabular}
\label{tab:ilp_groundaircraft_scip}
\end{table}

\begin{figure}[!h]
\centering
\includegraphics[scale=0.35]{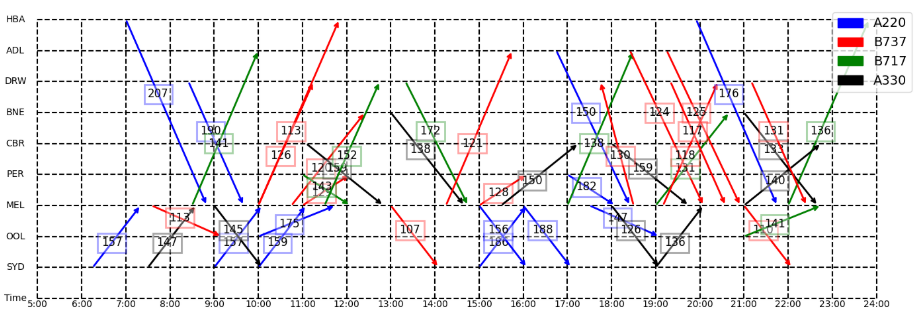}
\caption{Flight to Fleet assignment (ILP) Time-line network (SCIP) depicting the flights departing and arriving at a particular location. Each fleet type is depicted with a different color. Flights are assigned based on the objective of minimizing the cost, considering the difference between capacity and demand of passengers while meeting a set of operational and logistical constraints.  The incoming arrow shows the flights arriving at a location and the outgoing arrows show the flights departing from that location.}
\label{fig:ilp_assgnflight_scip}
\end{figure}

\begin{figure}[!h]
\centering
\includegraphics[scale=0.35]{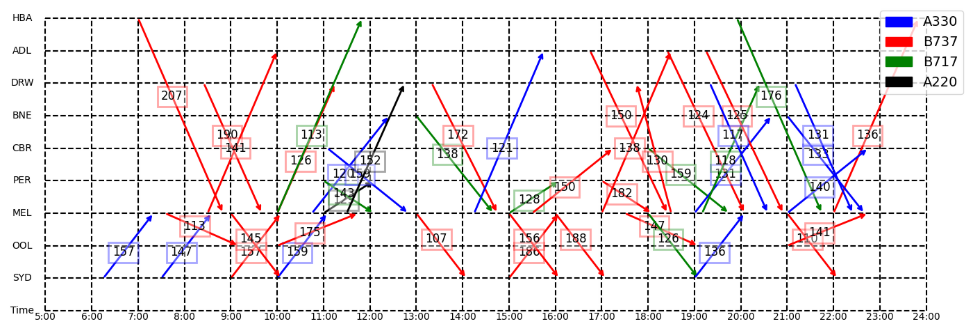}
\caption{Flight to Fleet assignment (ILP) Time-line network (LHCQM)}
\label{fig:ilp_assgnflight_cqm}
\end{figure}

\begin{figure}[!h]
\centering
\includegraphics[scale=0.55]{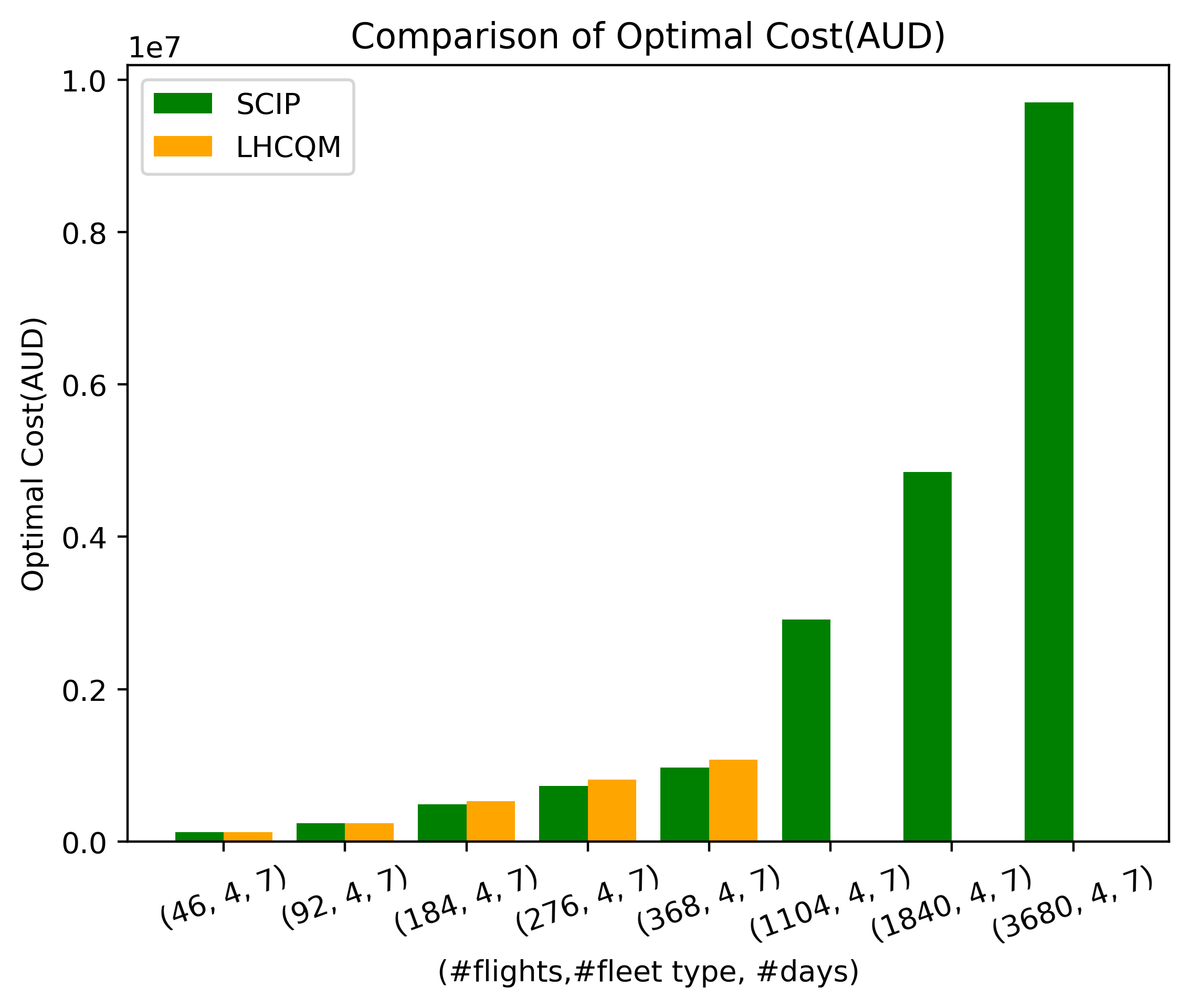}
\caption{Comparison of optimal cost (AUD) for both solvers in Fleet assignment (ILP) model}
\label{fig:ilp_optcost}
\end{figure}

\begin{figure}[!h]
\centering
\includegraphics[scale=0.55]{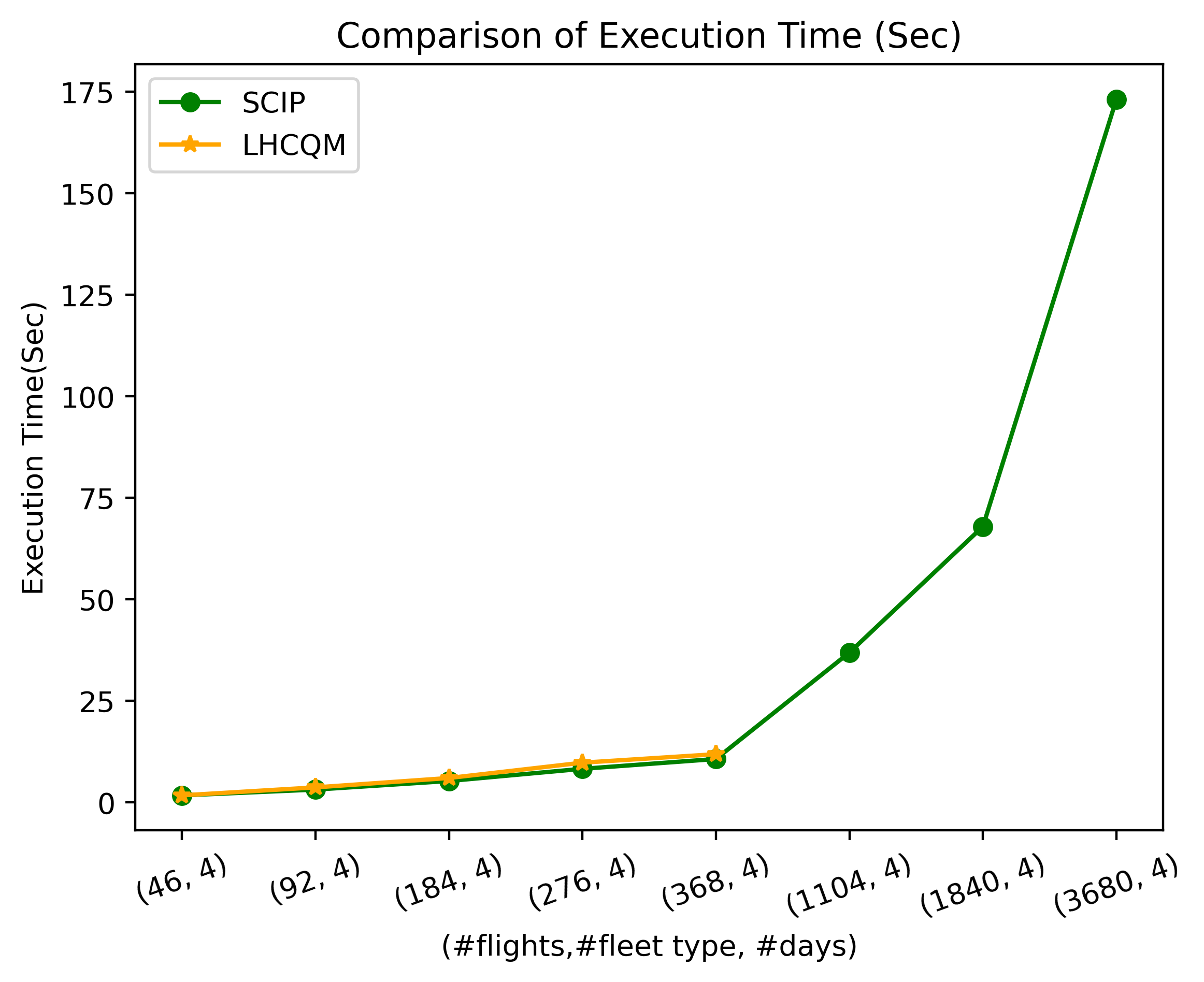}
\caption{Comparison of solution time (sec) for both solvers in Fleet assignment (ILP) model}
\label{fig:ilp_soltime}
\end{figure}

\newpage
\section{Conclusion}

The fleet assignment problem is a crucial optimization challenge that directly affects airline profitability and operational efficiency. By assigning the right aircraft types to flights, airlines can significantly reduce costs while maintaining service quality. In this study, we investigated two mathematical formulations of the Fleet Assignment Problem (FAP): a Binary Linear Programming (BLP) model and an Integer Linear Programming (ILP) model. We utilized both SCIP and Quantum Annealers (LHCQM) to solve instances of the problem, drawing on real-world data for our analysis. 

Our comparative analysis revealed that both SCIP and quantum annealers provide optimal solutions for smaller instances of the BLP model, but as the problem size increases, quantum annealers exhibit a clear advantage in terms of computational efficiency, delivering near-optimal solutions [Table \ref{tab:blp_optimalcost_exetime}] with reduced runtime [figure \ref{fig:blp_optcost}]. For the ILP model, however, quantum annealers struggle to solve larger instances, while SCIP remains more reliable, albeit with increased computational effort. The inclusion of the aircraft balance constraint in the ILP model adds further complexity, requiring advanced techniques to ensure feasible and optimal fleet utilization across multiple flight segments. 

This study highlights the potential of quantum annealing as a promising approach for large-scale optimization problems within the airline industry, particularly in instances where classical methods face scalability challenges. While quantum annealers show promise in solving certain problem sizes more efficiently, they still face limitations when tackling the more complex ILP formulations.

\section{Future Scope}
This research on the Fleet Assignment Problem (FAP) using both classical optimization methods and quantum computing opens up several promising avenues for future exploration. As quantum computing technologies advance, there is the potential for developing more efficient and scalable solutions to intricate airline scheduling challenges.Future research could focus on overcoming the infeasibility issues of the ilp model in the FAP. Future work could investigate the integration of fleet assignment with other scheduling problems, such as crew pairing, tail assignment, and aircraft routing. As quantum computing hardware improves, scalability remains a key challenge. Future studies could investigate the scalability of quantum algorithms in solving increasingly larger FAP instances and compare their performance against state-of-the-art classical solvers. Beyond airline operations, the methods developed for fleet assignment optimization can be extended to other industries, such as logistics, transportation, and supply chain management. Future research could explore the application of quantum computing and advanced optimization techniques to similar problems in these sectors, offering cross-industry benefits.     

\section{Acknowledgement}
Authors are grateful to Godfrey Claudin Mathais, C V Sridhar for their guidance.

\bibliographystyle{unsrt}
\bibliography{reference}

\begin{thebibliography}{10}

\bibitem{bib2}
Hanif~D. Sherali, Ebru~K. Bish, and Xiaomei Zhu.
\newblock Airline fleet assignment concepts, models, and algorithms.
\newblock {\em European Journal of Operational Research}, 172(1):1--30, 2006.

\bibitem{bib1}
Yavuz Ozdemir, Huseyin Basligil, and Kemal Nalbant.
\newblock Optimization of fleet assignment: A case study in turkey.
\newblock {\em An International Journal of Optimization and Control: Theories \&
  Applications (IJOCTA)}, 2, 01 2012.

\bibitem{bib18}
ZIB.
\newblock Scip, “scip web page".

\bibitem{bib13}
D-Wave~Systems Inc.
\newblock What is quantum annealing? — d-wave system documentation
  documentation.

\bibitem{bib12}
D-Wave~Systems Inc.
\newblock Constrained quadratic models — ocean documentation 8.0.0
  documentation.

\bibitem{bib3}
Barry~C Smith and Ellis~L Johnson.
\newblock Robust airline fleet assignment: Imposing station purity using
  station decomposition.
\newblock {\em Transportation Science}, 40(4):497--516, 2006.

\bibitem{bib4}
Russell~A Rushmeier and Spyridon~A Kontogiorgis.
\newblock Advances in the optimization of airline fleet assignment.
\newblock {\em Transportation science}, 31(2):159--169, 1997.

\bibitem{bib5}
Ram Gopalan and Kalyan~T Talluri.
\newblock Mathematical models in airline schedule planning: A survey.
\newblock {\em Annals of Operations Research}, 76(0):155--185, 1998.

\bibitem{bib6}
Jeph Abara.
\newblock Applying integer linear programming to the fleet assignment problem.
\newblock {\em Interfaces}, 19(4):20--28, 1989.

\bibitem{bib7}
Yana Ageeva.
\newblock {\em Approaches to incorporating robustness into airline scheduling}.
\newblock PhD thesis, Massachusetts Institute of Technology, 2000.

\bibitem{bib8}
Cynthia Barnhart, Timothy~S Kniker, and Manoj Lohatepanont.
\newblock Itinerary-based airline fleet assignment.
\newblock {\em Transportation Science}, 36(2):199--217, 2002.

\bibitem{bib9}
Catherine Mancel and Felix Mora-Camino.
\newblock Airline fleet assignment: a state of the art.
\newblock In {\em ATRS 2006, 10th air transportation research society
  conference}, 2006.

\bibitem{bib10}
Lu{\'\i}s~Noites Martins.
\newblock Applying quantum annealing to the tail assignment problem.
\newblock Master's thesis, Universidade do Porto (Portugal), 2020.

\bibitem{bib11}
Pontus Vikst{\aa}l, Mattias Gr{\"o}nkvist, Marika Svensson, Martin Andersson,
  G{\"o}ran Johansson, and Giulia Ferrini.
\newblock Applying the quantum approximate optimization algorithm to the
  tail-assignment problem.
\newblock {\em Physical Review Applied}, 14(3):034009, 2020.

\bibitem{bib15}
D-Wave~Systems Inc.
\newblock “d-wave systems web page,” . available: D-wave systems | the
  practical quantum computing company.

\bibitem{bib16}
D-Wave~Systems Inc.
\newblock Leap’s hybrid solvers, hybrid solvers — ocean documentation 8.0.0
  documentation.

\bibitem{bib19}
Amedeo Bertuzzi, Davide Ferrari, Antonio Manzalini, and Michele Amoretti.
\newblock Evaluation of quantum and hybrid solvers for combinatorial
  optimization.
\newblock In {\em Proceedings of the 21st ACM International Conference on
  Computing Frontiers}, pages 232--239, 2024.

\end{thebibliography}


\end{document}